\documentclass{egpubl}
\usepackage{eg2024}
 
\ConferenceSubmission

\usepackage[T1]{fontenc}
\usepackage{dfadobe}  

\usepackage{cite}  %
\usepackage{multirow}

\BibtexOrBiblatex
\electronicVersion
\PrintedOrElectronic
\ifpdf \usepackage[pdftex]{graphicx} \pdfcompresslevel=9
\else \usepackage[dvips]{graphicx} \fi

\usepackage{egweblnk}

\usepackage{amsmath,amssymb,amsfonts}%
\usepackage{xcolor}%
\usepackage{subfigure}
\usepackage{duckuments}
\usepackage{mathtools}
\usepackage{array}
\usepackage{siunitx}

\definecolor{rev}{rgb}{0,0,0}

\title[Utilizing Motion Matching with Deep Reinforcement Learning for Target Location Tasks]%
      {Utilizing Motion Matching with Deep Reinforcement Learning for Target Location Tasks}

\author[Jeongmin Lee \& Taesoo Kwon \& Hyunju Shin \& Yoonsang Lee]
{\parbox{\textwidth}{\centering Jeongmin Lee$^{1,2}$\orcid{0000-0002-7774-1427},
  Taesoo Kwon$^{2}$\orcid{0000-0002-9253-2156}, 
  Hyunju Shin$^{1}$\orcid{0009-0009-7367-4149} 
  and Yoonsang Lee$^{2}$\orcid{0000-0002-0579-5987}
  }
        \\
{\parbox{\textwidth}{\centering $^1$Samsung Electronics Co., Seoul R\&D Campus, South Korea\\
         $^2$Hanyang University, Department of Computer Science, South Korea
       }
}
}

\begin{document}

\maketitle
\begin{abstract}
We present an approach using deep reinforcement learning (DRL) to directly generate motion matching queries for long-term tasks, particularly targeting the reaching of specific locations.
    By integrating motion matching and DRL, our method demonstrates the rapid learning of policies for target location tasks within minutes on a standard desktop, employing a simple reward design.
    Additionally, we propose a unique hit reward and obstacle curriculum scheme to enhance policy learning in environments with moving obstacles.

\ccsdesc[300]{Computing methodologies~Motion processing}
\ccsdesc[300]{Computing methodologies~Motion path planning}

\printccsdesc   
\end{abstract}  

\section{Introduction}
Generating character animation in virtual environments has been a long challenge in the computer graphics society.
Among various approaches, motion matching~\cite{GDC_2016} is a widely known kinematic method, popular in game industry for its simplicity while still achieving a relatively high quality of motion.
The method extracts low-dimensional features from each posture and regularly searches for the next best fitting posture.
This greedy search aims to satisfy both smooth transitions and user goals simultaneously.

Utilizing interactive input devices like a gamepad or joystick, motion matching effectively provides immediate control and generates full-body character motion.
However, in cases where handling larger datasets or performing more extended planning tasks is necessary, simple motion matching alone is not sufficient.
To address this, recent studies propose various methods that integrate the motion matching algorithm with deep learning, whether by replacing the processing steps of motion matching with neural networks~\cite{Holden_2020}, adopting complex structures that demand relatively long periods of training for achieving long-term tasks~\cite{Cho_2021}, or utilizing a teacher-student framework to achieve various levels of responsiveness~\cite{lee_tcp_2021}.
\textcolor{rev}{However, few methods have been proposed that allow for learning the target location task in just a few minutes with a simple structure.}

In this paper, we present an approach to train a policy using deep reinforcement learning (DRL), enabling direct generation of motion matching queries for long-term tasks, particularly those related to reaching target locations.
By combining motion matching and DRL, we demonstrate that a policy for performing target location tasks can be quickly learned within a short timeframe (as little as a few minutes on a standard desktop) using a simple reward design.
Additionally, we propose a novel reward term and curriculum design to facilitate the learning of target location task policies in environments with moving obstacles.

\section{Related Work}

\textcolor{rev}{Researchers have proposed various methods to enhance and diversify motion matching.}
\cite{Holden_2020} proposed a method of improving the speed of the motion matching process and reducing memory usage by applying supervised learning to the internal processes of motion matching.
\cite{deepphase2022} presented the PAE (periodic autoencoder) for learning a low-dimensional phase manifold and demonstrated the generation of high-quality motions by using the phase vector in this manifold as the motion matching feature.
~\cite{jeongmin_2023} introduced a long-horizon motion matching (LHMM), which involves selecting the motion matching query capable of generating optimal motions when considering a time range longer than the typical future interval length used in motion matching queries.

\textcolor{rev}{DRL has been utilized for enabling character actions in simulations, performing tasks like dribbling with simple motion data~\cite{2017-TOG-deepLoco}, replicating motions for goals~\cite{peng_2018}, or moving without motion data~\cite{yu_learning_2018}.
Various approaches use or adapt motion data for actions across contexts~\cite{won_physics-based_2022,jump2021}, with studies on efficient learning for adaptable policies in multiple scenarios~\cite{kwon_adaptive_2023}.}

Among various studies, our work is most closely related to the following two studies.
\cite{Cho_2021} involves clustering discrete state and action spaces using VQ-VAE to generate character motions based on motion matching.
Policies are trained using Q-learning in the clustered space and various structures such as a passive action table and action candidate table are maintained for this.
In contrast, we propose a simpler structure and efficiently train the policy with PPO in a continuous space to achieve similar results.
\cite{lee_tcp_2021} employs the RL step with motion matching for state transition in training the teacher policy.
However, their objective is not to learn the teacher policy itself, but rather to utilize it to train a student policy capable of achieving goals with higher-quality motions within shorter response time limits.
In contrast, our method involves training a policy based on motion matching with a simple structure to achieve long-term goals and 
a dedicated reward design and curriculum learning scheme to better learn policies in obstacle avoidance environments.

\section{RL Formulation for Plane Environmnet}\label{reinforcement-learning}

\textcolor{rev}{Our policy network inputs state and goal, outputs an action that serves as a query for the motion matching stage, outputting the next motion frame every 6 frames.}
We employ the following two environments for the task of guiding the character to reach the target location.
In this section, we will describe the RL formulation for Plane environment, where the objective is for the character to reach the target location without any obstacles.

\textbf{Motion Matching.}
Our method is based on motion matching ~\cite{GDC_2016} with typical matching features for human locomotion. 
A feature at frame $i$, $\mathbf{f}_i = \{\mathbf{c}_i, \mathbf{t}_i\} \in \mathbb{R}^{27}$, is composed of the character's current pose feature $\mathbf{c}_i$ and its future trajectory feature $\mathbf{t}_i$ for the next one second.
We extract $\mathbf{c}_i  = \{ \mathbf p^\textrm{lfoot}_i, \mathbf p^\textrm{rfoot}_i, \mathbf{v}^\textrm{lfoot}_i, \mathbf{v}^\textrm{rfoot}_i, \mathbf{v}^\textrm{root}_i \} \in \mathbb{R}^{15}$,
where $\mathbf p$ and $\mathbf{v}$s are the positions and velocities of the left and right foot, and the root (pelvis), with respect to the character frame consisting of the root forward vector, global up vector, and their cross product and originating at the horizontal root position.
We extract $\mathbf{t}_i  = \{\tau_{i+10}, \mathbf{d}_{i+10}, \tau_{i+20}, \mathbf{d}_{i+20}, \tau_{i+30}, \mathbf{d}_{i+30}\} \in \mathbb{R}^{12} $,
where $\tau$ and $\mathbf{d}$ are the horizontal position and heading direction of the root, respectively.

Motion matching regularly seeks for the next motion frame $j$ that is closest to the query $\mathbf q$ ($j = \textrm{argmin}_k \| \mathbf{q} - \mathbf{f}_k \|^2$) and the character's motion is then updated by playing the frames that follow frame $j$ up to the next matching time point.  
In our formulation, this process of one matching and playback corresponds to one RL step.

\textbf{State} $\mathbf{s}_t$ is described as follows:
\begin{equation}
  \label{eq:plain-state}
  \mathbf{s}_t = \{\mathbf{c}_t, \mathbf{g}_t\},
\end{equation}
where $\mathbf{c}_t$ is the current pose feature at the RL step $t$ and $\mathbf{g}_t \in \mathbb{R}^{2}$ is the given horizontal target location with respect to the character frame.

\textbf{Action} $\mathbf{a}_t$ is described as follows:
\begin{equation}
  \label{eq:plain-action}
\mathbf{a}_t = \{\mathbf{t}_t \},
\end{equation}
where $\mathbf{t}_t$ is the future trajectory feature at $t$.
Note that $\mathbf a_t$ is a part of the motion matching query $\mathbf q_t$.
At each step $t$, the $\mathbf q_t$ is constructed by combining $\mathbf c_t$ in $\mathbf s_t$ and $\mathbf t_t$ in $\mathbf a_t$ and then the motion matching algorithm searches for the next closest frame.

\textbf{Reward} $r_t$ is described as follows:
\begin{equation}
  \label{eq:plain-reward}
    r_t = \mathrm{exp}(-\mathrm{dist}(\mathbf{s}_t)), \quad \textrm{where} \enspace \mathrm{dist}(\mathbf{s}_t) = \|\mathbf{g}_t\|.
\end{equation}
Note that $\mathrm{dist}(\cdot)$ represents a Euclidean distance between the horizontal root position and the target location because $\mathbf g_t$ is described with respect to the character frame.
Additionally, the agent is awarded one thousand rewards upon the successful completion of a episode (reaching the target).

\section{Extensions for Moving Obstacles Environment}\label{extensions-obstacles}

In this, we will explain the extensions for Moving Obstacle environment, where the character is required to reach the target location despite the presence of moving obstacles.

\begin{figure}
  \centering
    \includegraphics[trim=150 180 150 200, clip, width=0.46\columnwidth]{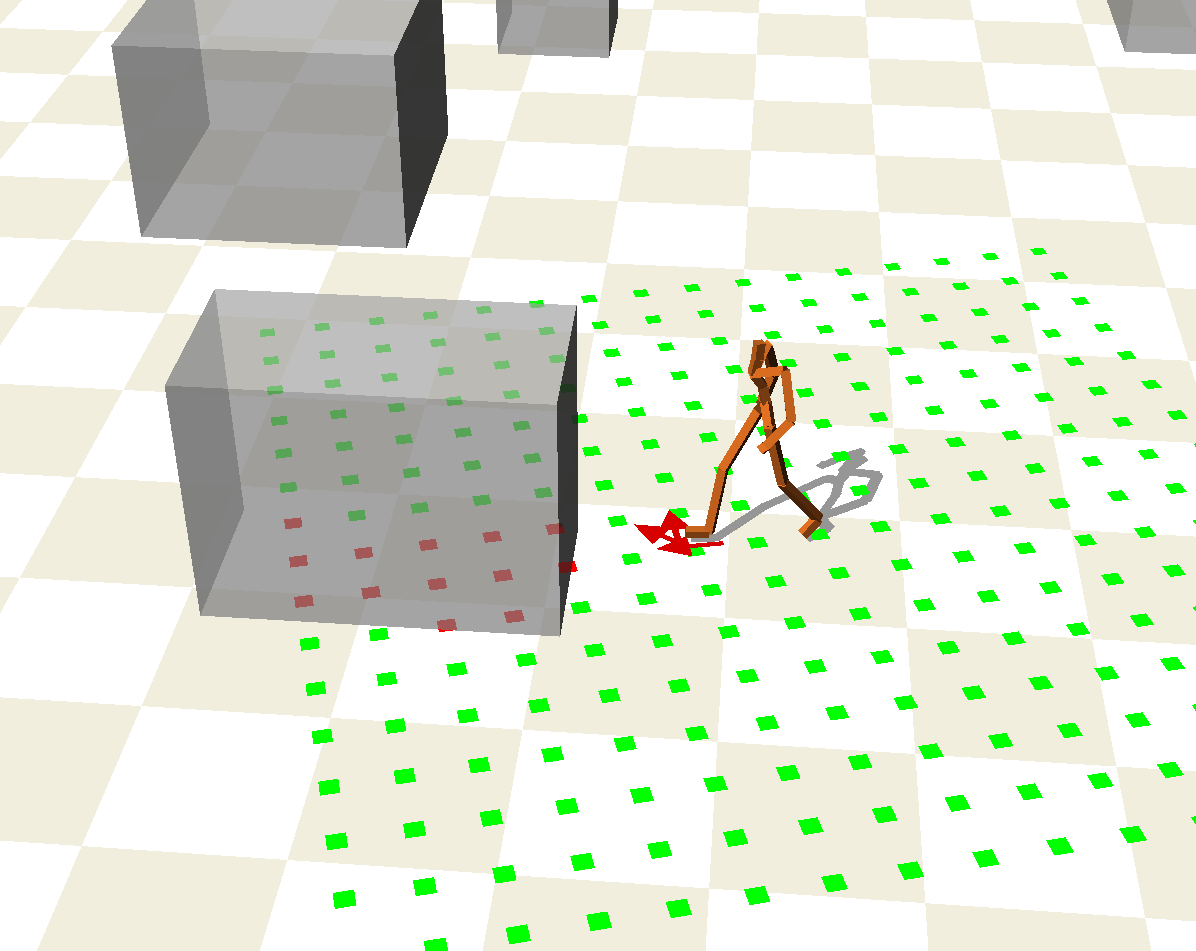}
    \includegraphics[trim=150 320 250 260, clip, width=0.46\columnwidth]{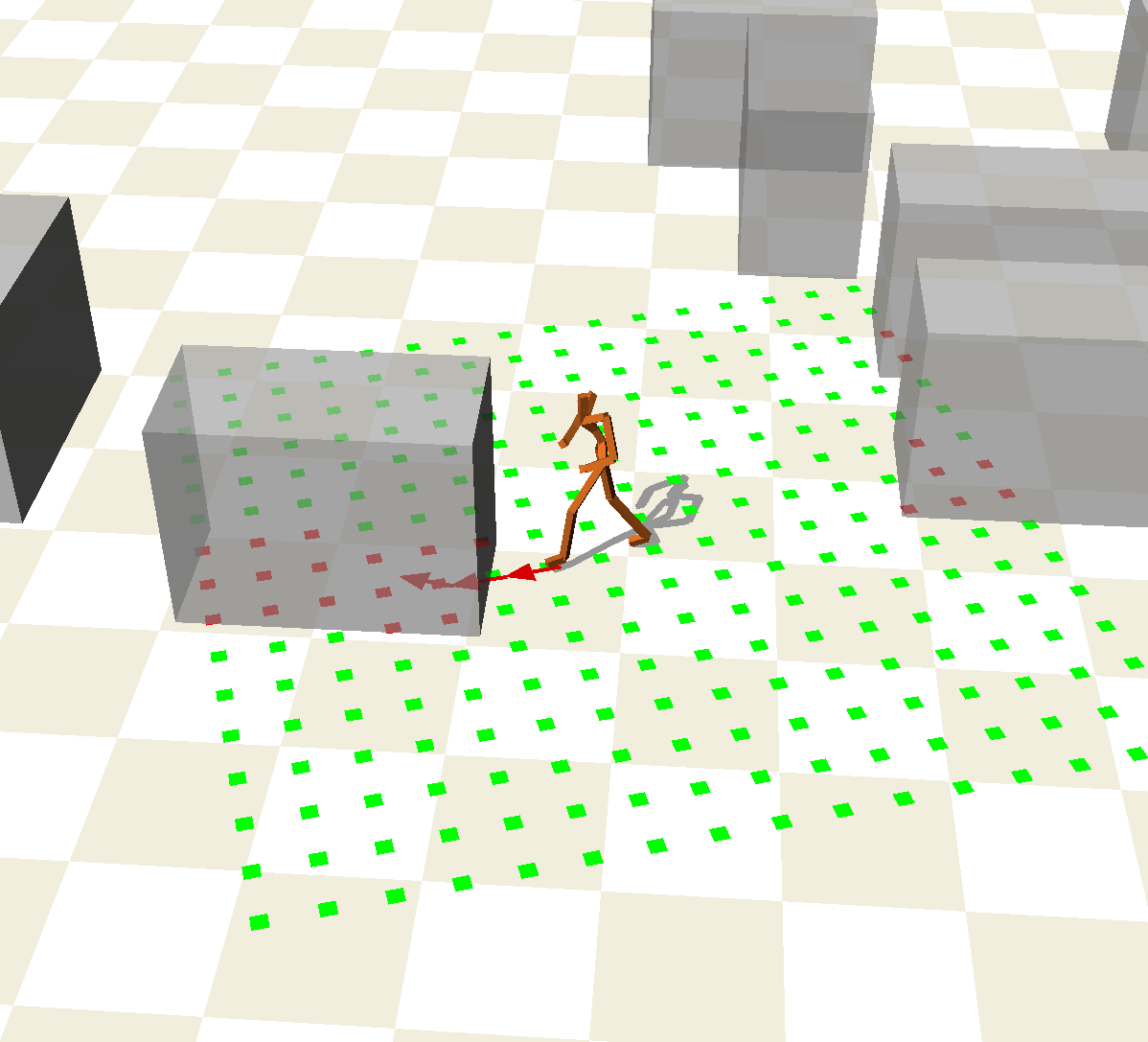}
  \caption{
      Examples of the hit reward.
       Left: The action (red arrows) results in a hit reward of $\mathrm{exp}(0)$ with no future positions in the obstacle.
       Right: The action leads to a hit reward of $\mathrm{exp}(-1)$ due to one future position inside the obstacle.
  }
  \label{fig:hits}
\end{figure}

\textbf{Hit Reward.} For Moving Obstacles environment, we introduce the additional reward term \textit{hit reward} that leverages the characteristics of our action design.
The total reward $r_t$ is described as follows:
\begin{equation}
  \label{eq:ext-reward}
    r_t = \mathrm{exp}(-\mathrm{dist}(\mathbf{s}_t)) + \mathrm{exp}(-\mathrm{hits}(\mathbf{a}_t)),
\end{equation}
where
\begin{align}
&\mathrm{hits}(\mathbf{a}_t) = \displaystyle\sum_{k=0}^{2}
\begin{cases}
1 &\text{if $\tau[k]$ is inside any obstacle} \\
0&\text{else}.
\end{cases}
\end{align}
The \textit{hit reward}, the second term in Equation~\ref{eq:ext-reward}, imposes a penalty on the count of future trajectory positions in an action that intersect with any obstacle (Figure~\ref{fig:hits}).
$\tau[k]$ is the $k$-th future horizontal root position in the action $\mathbf a_t$.
This term facilitates the effective learning of obstacle avoidance policies, by penalizing actions that would lead to collisions with obstacles within the next 1 second without actual execution of the future RL steps.
Additionally, the agent receives 10 rewards upon successfully completing an episode in this environment.

\textbf{Obstacle Curriculum.}
To enhance the learning of policies in environments with moving obstacles, we propose a curriculum learning scheme that gradually increases the sampling area for the target locations and the speed of the moving obstacles.

Specifically, in the initial stage of the curriculum, all obstacle speeds are set to 0 and the target locations are sampled within a $5 \textrm m \times 5 \textrm m$ rectangular area around the character's initial position.
In the last stage, the speeds are set to \SI{0.5}{m/s} and the sampling area expands to a $10 \textrm m \times 10 \textrm m$ rectangular area.
We incrementally increase the speeds of the obstacles and expand the sampling area for the target location as the stages progress in our 10-level curriculum scheme.
If the mean ratio of successful episodes (where the character reaches the target) for a policy exceeds 40\%, the curriculum advances to the next stage.

\begin{figure}
    \centering
    \includegraphics[trim=0 170 0 190, clip, width=0.9\columnwidth]{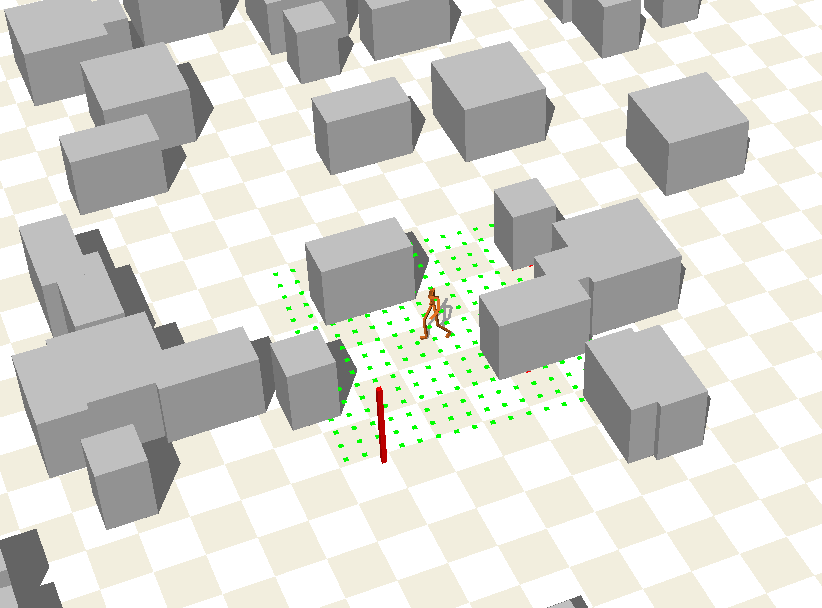}
    \caption{Example of the obstacle map.}
    \label{fig:sensor}
\end{figure}

\textbf{Additional Sensory Input.} In this environment, the agent takes an additional sensory input to detect nearby obstacles.
The state $\mathbf s_t$ is described as follows:
\begin{equation}
  \label{eq:ext-state}
  \mathbf{s}_t = \{\mathbf{c}_t, \mathbf{g}_t, \mathbf{o}_t\},
\end{equation}
where $\mathbf{o}_t \in \mathbb{R}^{16 \times 16 \times 2}$ is composed of two obstacle maps at the current and previous RL steps, \textcolor{rev}{similar to those used in \cite{2017-TOG-deepLoco},} each covering $6\textrm m \times 6\textrm m$ area (Figure~\ref{fig:sensor}).
Each map is generated from the readings of $16 \times 16$ binary sensors.
We collect two consecutive obstacle maps to capture the movement of obstacles.

\section{Training}\label{sec:training}

\begin{figure}
  \centering
    \includegraphics[trim=0 0 0 5, clip, width=0.36\columnwidth]{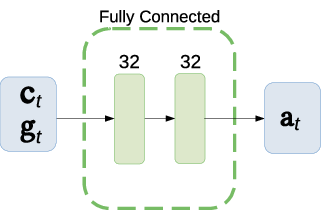}
    \hspace{.19cm}
    \includegraphics[trim=0 11 0 5, clip, width=0.6\columnwidth]{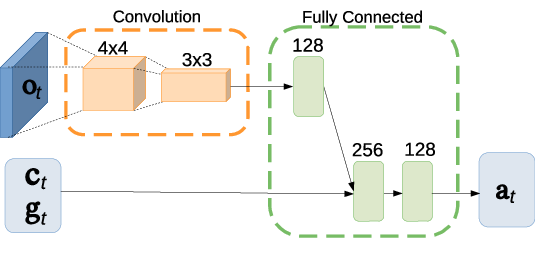}
  \caption{
      Our policy networks for Plain Environment (left) and Moving Obstacles Environment (right).
  }
  \label{fig:fc-network}
\end{figure}

\textbf{Plane Environment.} 
When learning policies in this setting, the target location $\mathbf{g}_t$ is sampled in a $10 \textrm m \times 10 \textrm m$ rectangular area around the character's position at the beginning of each episode.
Each episode starts with a character posture at the random frame in the motion dataset, which helps the agent to reduce redundant explorations.
An episode ends when the character comes within a \SI{0.5}{m} radius of the target location or surpasses the maximum step limit.

Our policy network for this environment consists of $32 \times 32$ FC layers (Figure~\ref{fig:fc-network}).
The value network follows the same structure, with the exception of having a single linear output unit.

\textbf{Moving Obstacles Environment.}
In this environments, 100 obstacles move in a maximum of \SI{0.5}{m/s}.
The obstacle sizes vary randomly, with a maximum dimension of $3 \textrm m \times 3 \textrm m$.
At the beginning of each episode, their initial positions are sampled in a $20 \textrm m \times 20 \textrm m$ rectangular area around the character.
An episode terminates whenever the character collides with an obstacle.
Unlike Plane environment, we do not impose a maximum step limit for an episode, and there are no extra rewards for successful completion.

Figure~\ref{fig:fc-network} illustrates the structure of our policy network for this environment.
The convolutional part uses 16 and 32 filters of $4 \times 4$ and $3 \times 3$ sizes with the stride of 1.
The output from the convolution layers is processed by a 128 FC layer, and then concatenated with $\mathbf c_t$ and $\mathbf g_t$  and processed by $256 \times 128$ FC layers.
Similarly, the value network follows the same structure, except the output unit.

\section{Experimental Results}\label{results}

In all experiments, we performed motion matching and policy learning based on it using the \textit{locomotion} dataset from \cite{jeongmin_2023} which is approximately 39 minutes long and comprises motions excluding jumps and t-poses from the dataset utilized in \cite{holden_phase_2017}.
One RL step corresponds to a motion progression of 0.2 seconds (equivalent to 6 motion frames in our \SI{30}{Hz} dataset), representing the interval between each motion matching query.
The motion matching process is followed by simple motion stitching to ensure smooth transitions, and analytic two-joint inverse kinematics to prevent foot sliding artifacts.

All the policies and experiments were trained and conducted on a i7-12700 processor with 12 cores and GeForce GTX 1650 GPU.
The policies were trained using the PPO implementation of RLLib~\cite{liang2018rllib}, with 11 rollout workers and a single trainer.
The animation results can be best observed in the accompanying video.

\begin{figure}
  \centering
  \subfigure[]{
    \includegraphics[trim=10 10 10 10, clip, width=0.46\columnwidth]{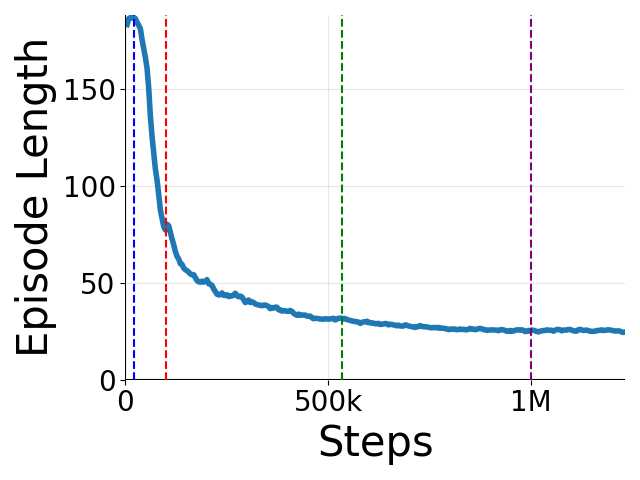}
  }
  \subfigure[]{
    \includegraphics[trim=10 10 10 10, clip, width=0.46\columnwidth]{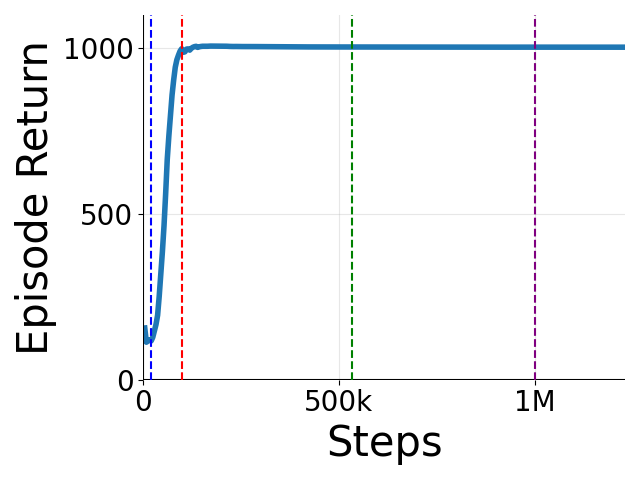}
  }
  \subfigure[]{
    \includegraphics[trim=10 170 150 200, clip, width=0.23\columnwidth]{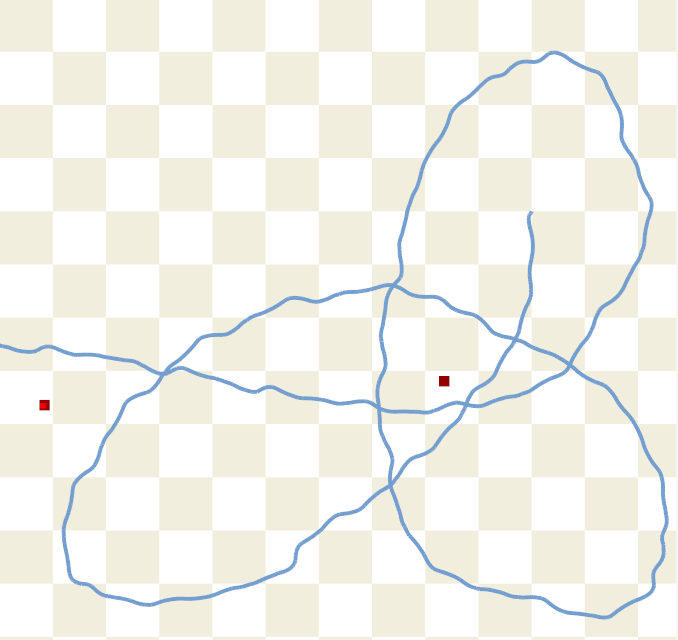}
    \includegraphics[trim=10 10 10 10, clip, width=0.23\columnwidth]{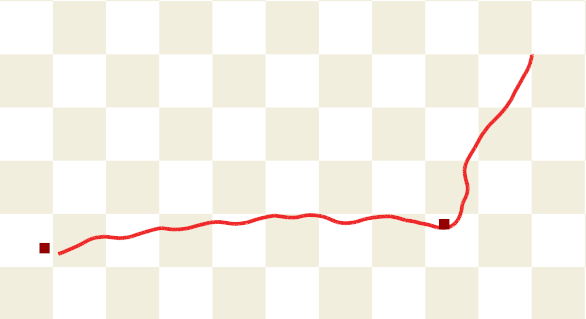}
    \includegraphics[trim=10 10 10 10, clip, width=0.23\columnwidth]{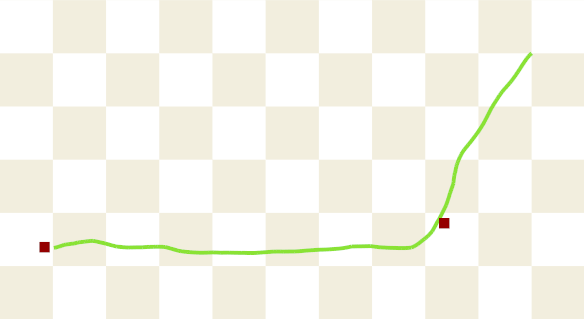}
    \includegraphics[trim=10 10 10 10, clip, width=0.23\columnwidth]{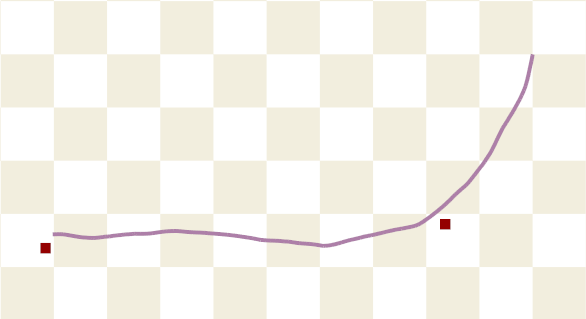}
  }
  \subfigure[]{
    \includegraphics[trim=60 110 80 170, clip, width=0.46\columnwidth]{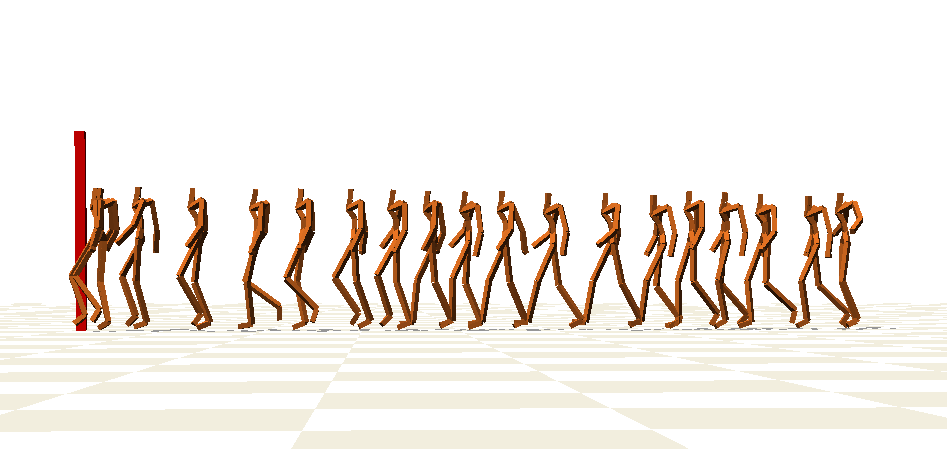}
  }
  \subfigure[]{
    \includegraphics[trim=60 100 80 140, clip, width=0.46\columnwidth]{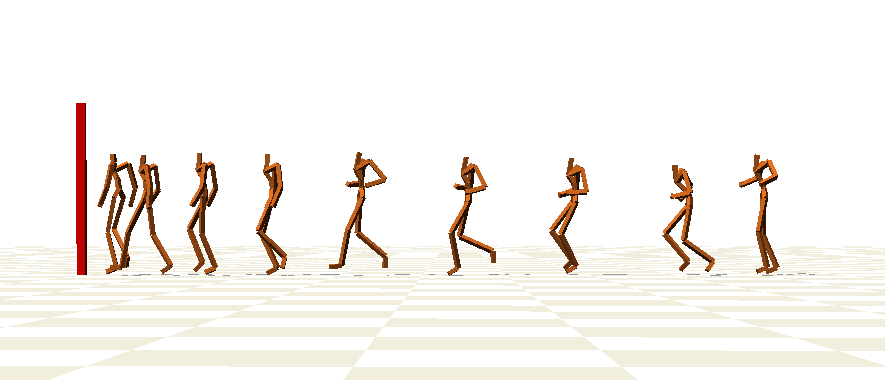}
  }
  \caption{
Learning curves for the plain environment. 
The blue, red, green, and purple vertical dashed lines in (a) and (b) correspond to the policy's performance at approximately 20k, 100k, 533k, and 1M steps, corresponding to 30, 150, 800, and 1500 seconds of training time.
The character's trajectory for each policy is illustrated in (c) using the corresponding color.
(d) and (e) depict the movement styles of policies trained for 100k and 1M steps, respectively.
  }
  \label{fig:plane-env}
\end{figure}

\textbf{Performance in Plane Environment.}
Our policy achieves the goal of reaching the target location with only a small number of samples and short training periods as the generation of full-body motion is based on motion matching, obviating the necessity for the policy to learn full-body motion generation.

Even with a training time as short as 30 seconds, a policy that reaches the target location can be achieved. However, in such cases, the character tends to experience significant delays when changing direction towards a different target.
Interestingly, as can be seen in Figure~\ref{fig:plane-env}, with the progression of training, the character exhibits a quicker change of direction and starts moving in a running motion to reach the target location more swiftly.

\begin{figure}
  \centering
  \subfigure[]{
    \includegraphics[trim=10 10 10 10, clip, width=0.46\columnwidth]{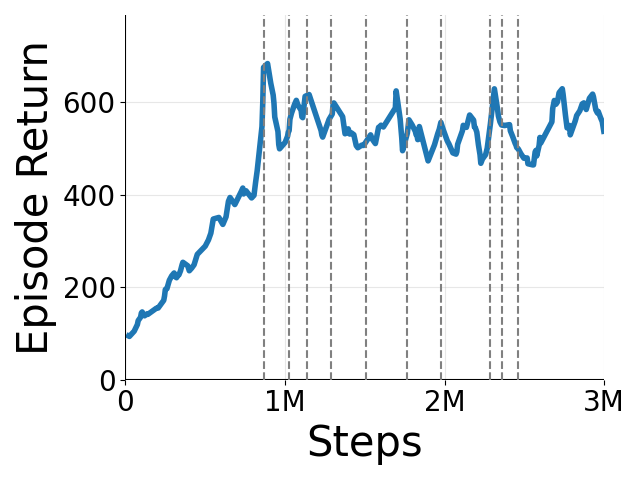}
  }
  \subfigure[]{
    \includegraphics[trim=10 10 10 10, clip, width=0.46\columnwidth]{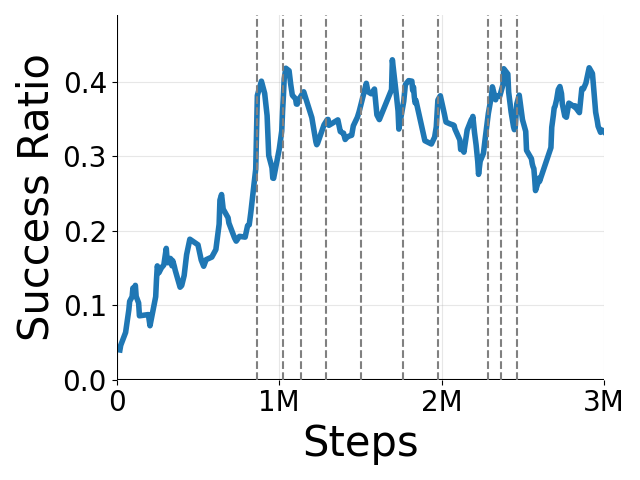}
  }
  \caption{
Learning curves for the moving obstacles environment in the early stages (before 3M steps).
The dashed lines signify the points at which the curriculum stage transitions to the next stage.
}
  \label{fig:moving-obstacles}
\end{figure}

\textbf{Performance in Moving Obstacles Environment.}
As depicted in Figure~\ref{fig:moving-obstacles}, our policy successfully traversed through all the 10 stages of the \textit{obstacle curriculum}.
The most recent progress occurred after the 492nd policy update, approximately at the 2.4 millionth step.
The final policy was obtained after training for a total of 14M steps over 8 hours.

\begin{figure}
  \centering
  \subfigure[]{
    \includegraphics[trim=0 10 0 0, clip, width=0.46\columnwidth]{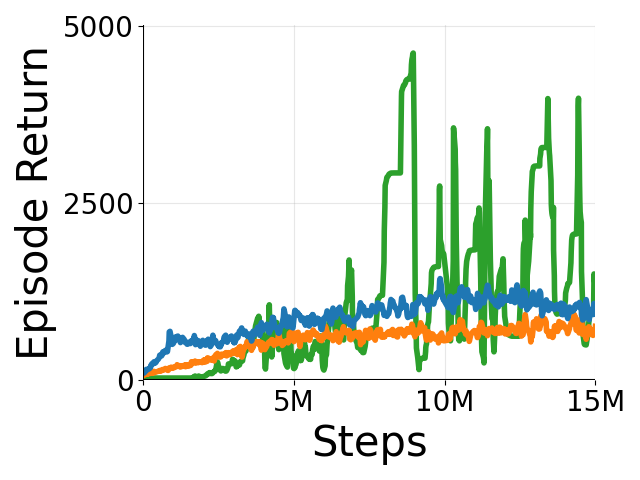}
  }
  \subfigure[]{
    \includegraphics[trim=0 10 0 0, clip, width=0.46\columnwidth]{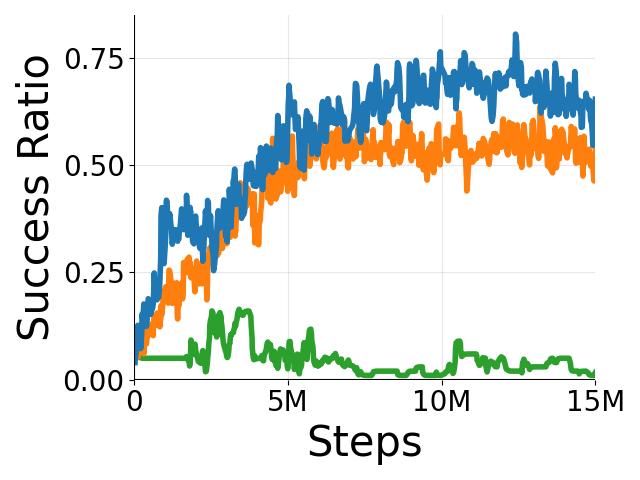}
  }
  \caption{
    Learning curves for the ablation study for the moving obstacles environment. Blue: ours. Green: without hit reward. Orange: without obstacle curriculum.
  }
  \label{fig:obstacle-ablation}
\end{figure}

\textbf{Ablation Study for Moving Obstacles Environment.}
We conducted an ablation study on each component of our extensions for Moving Obstacles environment.
Initially, we attempted to learn obstacle avoidance without the \textit{hit reward}. 
Subsequently, we explored learning without the \textit{obstacle curriculum}. 
In this scenario, the environment is set to the most challenging stage from the start.

Figure~\ref{fig:obstacle-ablation} illustrates the learning curves.
Our method demonstrates stable episode lengths and records the highest mean success ratio.
The ablation of the \textit{obstacle curriculum} converges to a lower success ratio and episode returns.
Notably, the ablation of the \textit{hit reward} results in highly unstable learning and significantly low success ratio.

\section{Discussion}\label{discussion}
In this paper, we introduce an approach employing DRL to directly generate motion matching queries for long-term tasks, with a specific focus on reaching target locations.
We observed a notable improvement in learning target location tasks in environments with moving obstacles through the proposed \textit{hit reward} and \textit{obstacle curriculum} scheme.

Our method, being based on motion matching, has limitations of high runtime memory usage and slow exploration speed.
These constraints could potentially be addressed by applying the methods proposed in \cite{Holden_2020}.
Additionally, the utilization of autoencoders, as demonstrated in studies such as \cite{deepphase2022}, for feature compositions holds the potential to enable various applications across diverse datasets and tasks.

\section*{Acknowledgements}
This work was supported by National Research Foundation of Korea (NRF) grant funded by Korea Government (MSIT) (RS-2023-00222776), with Yoonsang Lee and Hyunju Shin as corresponding authors.

\bibliographystyle{eg-alpha-doi} 
\bibliography{egbib}
\end{document}